# Magnetic nanofibers for remotely triggered catalytic activity applied to the degradation of organic pollutants

J.A. Fuentes-García[1*], B. Sanz[2], R. Mallada[1,3], M.R. Ibarra[1,4], G. F. Goya[1,4]

[1] *Instituto de Nanociencia y Materiales de Aragón (INMA), CSIC-Universidad de Zaragoza, 50018 Zaragoza, Spain.*

[2] nanoScale Biomagnetics S.L., 50197, Zaragoza, Spain

[3] Department of Chemical & Environmental Engineering, University of Zaragoza, 50018 Zaragoza, Spain

[4] *Departamento de Física de la Materia Condensada, Facultad de Ciencias, Universidad de Zaragoza, 50009 Zaragoza, Spain.*

*Corresponding author: jesus_spirit69@hotmail.com, j.fuentes@unizar.es*

ORCID of authors:

J.A. Fuentes-García: **0000-0003-4952-3702**

B. Sanz: **0000-0002-5578-7635**

R. Mallada: **0000-0002-4758-9380**

M.R. Ibarra: **0000-0003-0681-8260**

G. F. Goya: **0000-0003-1558-9279**

## Highlights

- New electrospun polymer nanofibers with embedded magnetic nanoparticles were assembled to produce magnetic composite nanofibers and build hydrophobic membranes.

- These magnetic nanofibers, remotely activated by alternate magnetic fields for inductive heating, were capable to triggering the degradation of the organic dye methylene blue as a proof of concept.

- The magnetic nanofibers were remarkably stable against extreme changes in pH and also to successive heating cycles, retaining their full catalytic performance.

- Degradation efficiencies >80% of methylene blue organic dye were obtained, showing that scaling-up for environmental applications is potentially feasible.

# TOC

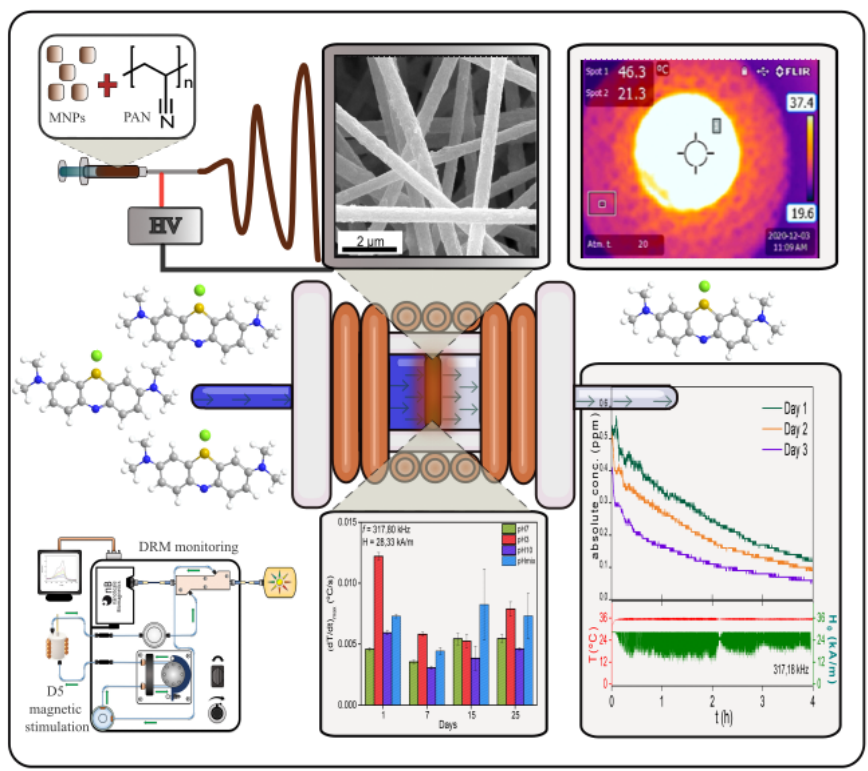

## ABSTRACT

This work reports on the synthesis and characterization of a new type of electrospun magnetic nanofibers (MNFs), and their application for degradation of organic pollutants using remote magnetic inductive heating. We describe a simple protocol combining a fast (≈5 min) synthesis of $MnFe_2O_4$ magnetic nanoparticles (MNPs) by sonochemical route, optimized for inductive heating, with their subsequent incorporation in electrospun MNFs composed of polyacrylonitrile (PAN) nanofibers. The resulting multifunctional MNFs (average diameter $\phi = 760 \pm 150$ nm) contain up to ≈30%wt. of the MNPs. The composite showed superhydrophobic behaviour ($\theta_c = 165º$) and a band gap value of 1.75 eV. We found that the presence of MNPs embedded into the polymeric nanofibers modify the exothermic and the glass transitions temperatures compared with pure PAN nanofibers, suggesting a strong attachment between MNPs and polymeric chains. The MNFs could be remotely activated by alternating magnetic fields (AMF, $f$ = 200-800 kHz, $H_0$=10-36 kA/m) for accelerating the catalytic reactions of the organic dye methylene blue (MB). A remarkable stability of the MNFs against degradation under extreme pH conditions (3 < pH < 10) resulted in a sustained heating efficiency after many heating cycles. We observed a degradation efficiency > 80 % in the presence of hydrogen peroxide under AMFs, attributed to $Fe^{2+/3+}$ and $Mn^{2+/3+/4+}$ active centers on the surface of the MNP/MNFs observed from XPS data. The capacity of these materials for magnetic remote activation appeals catalytic applications under conditions of darkness or restrained access, where no photocatalytic reactions can be achieved.

# 1. INTRODUCTION

Polymer-based nanofibers are a common choice for the design of novel multifunctional materials in both basic research and industrial sectors.[1] Current applications of these materials cover a wide spectrum including medical smart drug delivery systems [2], tissue engineering scaffolds [3], active wound dressings [4], electrodes for batteries [5], supercapacitors [6], and fuel cells [7]. Designing optimized polymer-based fibers requires a fine control of their physicochemical properties such as composition, diameter, orientation, mechanical performance, or biocompatibility. [8] [9] [10]. Their uses as water remediation materials for devices has also attracted much attention during the last years. [11-14]

For the fabrication of polymeric nanofibers, the electrospinning technique provides a simple and unexpensive option potentially scalable to the industrial manufacturing levels.[15] The flexibility of the electrospinning technique to produce a variety of tailored nanofibers relays on the accurate control of the main physical parameters during fabrication, such as the applied voltage or the nozzle-collector working distance, the injection rates and the viscoelastic properties of the precursor solution. [16, 17]

The synthesis of composite nanofibers, e.g., by embedding nanoparticles within the polymeric base material, has been successful to enhance optical, mechanical, or magnetic properties of the active materials.[18-20] Adding magnetic nanoparticles to the precursor polymers results in magnetic nanofibers that can be remotely actuated by magnetic fields in different ways, included magnetic heating applications. [21-24] Optimizing magnetic nanoparticles (MNPs) for heating uses requires controlling MNP's size, shape and composition to maximize the specific loss power (SLP, i.e., the power losses per unit mass of magnetic material). Among the large variety of synthesis routes to produce MNPs, the sonochemistry in aqueous media has been proven to be a fast, cheap, and green method that can provide some grams of magnetic material in few minutes of reaction.[25, 26]. Physical and chemical effects produced by the interaction of ultrasound irradiation within aqueous solutions provide the conditions for fast crystallization and controlled growth of nanostructures. [27, 28] We have previously reported a systematic screening of the physical parameters such as irradiation power and concentration of precursors, time, and temperature, to obtain optimized MNPs for heating purposes. [29] The combination of both electrospinning and sonochemistry to produce new hybrid systems provides a promissory way to obtain multifunctional materials using this simple approach with minimal infrastructure.

Hybrid MNP-embedded polymer nanofibers can be used for the degradation of organic pollutants from water [30], which can be oxidized and/or degraded by redox catalysis thanks to the presence of transition metal ions onto the MNP´s surfaces [31]. Transition metal ions are efficient mediators for electron transfer with other ions and molecules in the Fenton-like cyclic reactions [32] in presence of hydrogen peroxide, producing hydroxyl radicals (OH˙) and other reactive oxygen species (ROS) that initiate radical polymerization, oxidation, or chlorination reactions [33, 34]. For these reactions, pH and temperature have strong influence their efficiency as degradation agents. [35]

In this context, the hybrid magnetic nanofibers (MNFs) provide a double purpose material: while the embedded MNPs optimized for inductive heating under AMF allows to increase the local temperature and its catalytic activity is promising, the polymeric nanofiber matrix support provides an inert and flexible platform with the ability to shape devices for different applications. Therefore, embedding the MNPs within nanofibers is an ideal strategy for their immobilization, avoiding the migration and loses of magnetic material from the matrix, allowing to recycle the catalyst after many cycles of pollutant degradation and preserving the systems efficiency. [30, 36]

In this work, we present a study on electrospun magnetic nanofibers (MNF) aimed as a remotely activated catalytic system, composed of a stable polyacrylonitrile (PAN) $\phi = 760 \pm 150$ nm nanofiber matrix to host iron-manganese oxide ($MnFe_2O_4$) MNPs. These MNPs were selected for a double purpose to act as remotely activated nanoheaters and as reaction center-points for degradation of heterocyclic organic tetra-methylthionine chloride dye (methylene blue MB), chosen as a proof-of-concept molecule. This approach constitutes an option where the sun radiation (photocatalysis) and ultrasound irradiation (sonocatalysis) cannot be applied to clean water wastes, for example in cases of underground pipes or small tubing.

## 2. EXPERIMENTAL SECTION

The chemicals and solvents used in this work were: iron (II) sulphate heptahydrate ($FeSO_4 \cdot 7H_2O$) ReagentPlus® ≥ 99%, manganese (II) sulphate monohydrate ReagentPlus® ≥ 99% ($MnSO_4 \cdot H_2O$), sodium hydroxide pellets (NaOH) ACS reagent ≥ 97 %, polyacrylonitrile (PAN, $(C_3H_3N)_n$) average Mw 150,000, N,N-dimethylformamide (DMF, $CHON(CH_3)_2$) ReagentPlus® ≥99% were purchased from Merck. Milli-Q deionized water 18.2 MΩ·cm was employed during sample preparation. Hydrochloric acid (HCl 37%, Labbox, Spain), nitric acid ($HNO_3$, 65%, Panreac) and ammonium thiocyanate ($NH_4SCN$, Merck) were employed for iron contents determination.

Irradiation power (specific energy input 0.27 J/mL, 20 kHz) was applied using an Ultrasonic Vibra-cell® VCX 130 equipment with Ti-6A1-4V 6 mm tip, immersed in a 100 mL borosilicate glass bottle ($\phi$ = 56, h = 100 mm) for sonochemical synthesis of magnetic nanoparticles. The electrospinning system employed for nanofiber processing was a laboratory-made version of a classical setup. A LEGATO® 101 syringe pump (Kd Scientific) was used to control the rate of polymer injection through an AISI 316 stainless steel needle (O$\phi$ = 1 mm, h = 20 mm). The needle was connected to the positive pole of an adjustable high voltage Power Supply Unit (Genvolt®, 30 W, 200 V - 30 kV), while the negative pole was connected to a circular metallic collector ($\phi$ = 100 mm, h = 30 mm) with adjustable displacement in z axis.

MNP were obtained mixing 8:2 ratio of $FeSO_4$ and $MnSO_4$ respectively in 1 L of Milli-Q water for a 5.4 mM solution. After, 90 mL of the $Fe^{+2}$ and $Mn^{+2}$ solution was ultrasonicated during 10 minutes to reach temperature T≈ 60 ºC. Then, adding 10 mL of NaOH solution (2N) to alkaline conditions for adequate MNP crystallization. The reaction is maintained another 10 minutes under continuous irradiation and after, the ultrasonic processor was turned-off. The resultant solution in the reactor was poured in a beaker with 100 mL of water for lower temperature and pH in the suspension. The described procedure was performed 10 times and the obtained MNP were washed with Milli-Q water using magnetic decantation until pH = 7 in the dispersion. After, we obtain the magnetic powder drying MNP (air atmosphere at 60ºC for three days). Using an ultrasonic bath for 10 minutes, 1 g of MNP powder was re-dispersed in 10 mL of DMF. The obtained dispersion was placed on magnetic stirring (250 RPM), and heated at 80 ºC to improve the solubility of the polymer. Then, 1 g of PAN was dissolved in the DMF-MNP dispersion for 2 hours. Heating was turned-off and stirring continued for 12 hours in order to homogeneous dispersion of MNP in the PAN matrix and leading to adequate entanglement of polymeric chains.

Polymeric films composed by PAN and MNP (MNFs samples) were processed injecting 1 mL of the previously prepared PAN-MNP solution within the electrospinning system. The feed rate was 0.7 mL/h, while distance needle-collector and high-voltage applied were 15 cm and 8 kV respectively. The magnetic membrane deposited on aluminum foil was trimmed in 10 mm diameter circles for further characterization and evaluation of their hyperthermia response and degradation capacity.

## 2.1 Characterization

The compositional characterization of the MNFs was performed in a CSEM-FEG Inspect™ F50 Scanning Electron Microscope using EDS analysis. Cross sections of the MNFs were done using a Focused Ion Beam FIB-SEM (Nova 200, FEI). All samples were coated with a ≈20 nm thickness carbon layer before observation. From image processing, average diameter and standard deviation were obtained. EDS analysis composition tables obtained allowed to estimate the sample stoichiometry (at %) of the $Fe_{3-x}Mn_xO_4$ samples (EDS composition tables are provided in the Supplementary Information File). A blank of PAN fibers (PANFs) was prepared for comparison with the MNFs using the following techniques. Normal modes of vibration from functional groups in the samples were analyzed using Attenuated Total Reflectance Fourier Transform Infrared spectroscopy (ATR FT-IR), the spectra from 4000 to 600 cm$^{-1}$ were obtained using Bruker VERTEX 70v FT-IR Spectrometer. Thermal analysis of PANFs and MNFs was performed using 1.5 mg of each sample for thermogravimetric analysis (TGA) in a Mettler Toledo TGA SDTA851 analyzer from 50 to 800 ºC, heating rate of 10 ºC min$^{-1}$, $N_2$ purge of 60 mL min$^{-1}$ in ceramic pan. Also, 1.5 mg of each sample was employed for Differential Scanning Calorimetry (DSC) in a DSC822e Module (Mettler Toledo) from 50 to 500 ºC (1ºC/minute) under $N_2$ atmosphere. The UV-vis spectra of solid membranes composed by PANFs and MNFs (thickness 30 µm) were performed in a JASCO V-670 UV-vis/NIR spectrophotometer (JASCO, Tokyo, Japan) using the solid-state diffuse reflectance technique in a 60 mm UV-vis/NIR with an integrating sphere from 200 nm to 800 nm, scanning step of 10 nm s$^{-1}$. The optical band-gap energy ($E_g$) was determined from the reflectance spectra, Tauc´s using plots $F^2$ vs. $hv$, where **F** is the Kubelka-Munk function of the reflectance, **h** is the Plank constant and **v** the frequency. [37] Magnetization curves as a function of temperature and applied field were obtained using a SQUID magnetometer (MPMS XL, Quantum Design) in the range -10 k Oe ≤ H ≤ 10 k Oe and temperatures within 10 K ≤ T ≤ 300 K. Conditioned gelatin capsules were filled with ≈1 mg of as-prepared MNFs for these measurements.

## 2.2 Magnetic heating performance

The specific loss power (SLP) of MNPs and MNFs was measured using the calorimetric method of heating curves vs. time. An alternating magnetic field generator with amplitudes $0 \leq H_0 \leq 71.2$ kA·m$^{-1}$ and frequencies 200 kHz ≤ f ≤ 800 kHz (nB nanoScale Biomagnetics, Spain) was employed. The sample was placed into a semi-adiabatic holder that uses a turbomolecular pump to dynamically produce low pressure (10$^{-6}$ mbar) within double wall jacket providing thermal insulation during the experiments (Figure 1 a)).

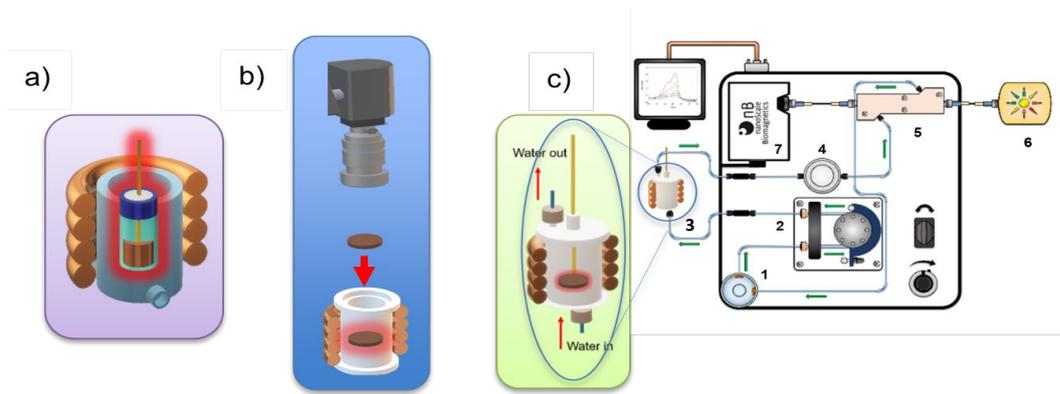

**Figure 1** Schematic illustration of the coil sets for inductive heating measurements: a) Calorimetric measurements within a thermally-insulated double wall jacket; b) non-adiabatic setup for thermographic measurements, placing the MNPs samples within the AMF and a zenithal thermographic camera; c) setup for dynamic conditions: 1. Water reservoir, 2. Peristaltic pump for controlled flux, 3. Coil, 4. Bubble trap, 5. Optical path for UV-Vis sampling, 6. UV-Vis lamp, 7. UV-Vis detector with interface for collecting data.

For the SLP measurements, 3 mg of MNPs were dispersed in water with a 10 %wt. commercial gelatin (Royal™) at 80 °C. After subsequent cooling the gelatin formed a consistent but workable matrix with melting temperatures of about 45 °C. This is a two-fold strategy intended to a) hinder MNPs precipitation or agglomeration during heating experiments and b) eliminate the contribution of Néel relaxation of the particles to the SLP, which is obtained from the heating curves using the following equation:

$$SLP = \frac{\rho \cdot c_l}{\varphi} \cdot \left(\frac{dT}{dt}\right)_{max} \qquad (Eq.\ 1)$$

where $\rho$ and $c_l$ are the density (1350 kg·m$^{-3}$) of the medium (gelatin) and the experimentally measured specific heat 4.1 kJ·kg$^{-1}$·°C$^{-1}$ [38]. The heating rate $\left(\frac{dT}{dt}\right)_{max}$ was extracted from the heating curves (T *vs* t) using a linear fit of the initial Dt ≈20 s heating where the process is approximately adiabatic. The SLP of the MNFs was measured in aqueous media, by placing 3 mg of *as prepared* MNFs in Milli-Q water (0.5 mL) and using the following expression:

$$SLP = \frac{m_{fiber}\, c_{fiber} + m_l\, c_l}{m_{fiber}} \cdot \left(\frac{dT}{dt}\right)_{max} \qquad (Eq.\ 2)$$

with $c_{fiber}$ = 0.46 J·kg$^{-1}$·°C$^{-1}$ is the specific heat of the PAN polymer [39], and $c_l$= 4182 J·kg$^{-1}$·°C$^{-1}$ is the specific heat of the liquid (water). $m_{fiber}$ and $m_l$ are the masses of the MNFs and water, respectively.

To obtain the SLP values, the concentration of MNPs in all samples was determined through colorimetric method, based in the determination of Fe contents using a UV-vis spectrophotometer (Shimadzu UV-128, Japan). For that, the total contents of the samples were dissolved in HCl 6 M and $HNO_3$ (65%) at room temperature to obtain a $Fe^{3+}$ ions solution. After, potassium thiocyanate was added to the $Fe^{3+}$ solution, forming a red color complex with absorbance peak at λ=478 nm. These band was monitored and compared with a calibration curve to quantitative analysis of the iron contents. Finally, the molecular weight of the magnetic materials was calculated using the obtained iron contents and the EDS data (Figure S1).

The magnetically induced heating on MNFs ($f$ = 317.80 kHz, $H_0$ = 28.33 kA·m$^{-1}$) was measured by placing a single circular cut of the MNFs in the sample holder schematized in Figure 1 b), and acquiring thermographic images with a thermal camera (FLIR F50 System, USA) at the zenithal position over the sample.

The heating capacity was evaluated at dynamic conditions of water flux under different AMF values using a DRM accessory equipment (nB nanoScale Biomagnetics, Spain) schematized in Figure 1 c). This setup has an internal chamber to pump any required liquid flux across the MNFs membranes. In our case, different solutions with acid, neutral, and basic pH, as well as a MB solution (see below), were employed.

The stability of *as-prepared* MNFs samples against pH was evaluated by pumping three different water solutions with pH=3, 7 and 10 across the MNFs membrane, under an AMF of $f$ = 317.80 kHz and H = 28.3 kA·m$^{-1}$ and measuring the evolution of the heating profiles. The experimental setup allowed us to modify *in situ* the media conditions by switching circuits with the different flowing solutions with changing pH during AMF heating of the MNFs samples. The experiments were designed to turn on the magnetic field for a specific time ($AMF_{on}$ = 10 min) and then running it off for $AMF_{off}$ = 20 minutes allowing the cooling of the samples. Three cycles of heating/cooling under different pH were produced in this way: the first with a pH = 7 solution, the second cycle a pH = 3 solution and the third cycle with a pH = 10 solution. This experiment is identified as 'pH mix'.

In all these experiments, each sample was measured in triplicate and sequentially four times in a month (1st, 7th, 15th and 25th days) to evaluate the reusability and reproducibility of the heating capacity of MNFs. Between experiments, the MNFs samples were stored at 60 ºC without washing, only pH mix sample was washed with a pH = 7 solution before storing.

## 2.3 Degradation kinetics of MB under magnetic stimulus

For the characterization of MB degradation under AMF stimulation in continuous mode, the setup was the same already describer in Section 2.2, but changing the composition of the circulating fluid by a dye-containing liquid that was to be degraded. As before the flow past the modular UV-Vis spectrophotometer allowed for quantitative determination of the dye concentration. A flow rate of 0.4 mL·min$^{-1}$ was set in a closed-circuit configuration, using a MB solution that flowed through the sample holder containing the MNFs sample (Figure 1 c) 3), placed within the magnetic coil. For these experiments, we selected a target temperature for the catalytic reaction, measured by an optical fiber thermometer *in situ,* which also provided the control of this target temperature during experiments through a feedback PI system to regulate the magnetic field amplitude and stabilize the temperature setpoint during the inductive heating of the MNFs.

The concentration of MB in real time was measured by a compact modular UV-vis spectrophotometer (USB2000+VIS-NIR) with an halogen light source (HL-2000-FHSA) (Ocean Optics™) with a bandwidth from 349 to 1019 nm. The flowing MB solution passed through a PEEK-flow cell (See element (5) in Figure 1c) with an optical path length of 100 mm and internal volume of 60 μL thus the real time absorbance was monitored. The unavoidable formation of micro-bubbles within the fluing solutions interfered with precise absorbance measurements, and thus a bubble tramp (4) was added before the flow cell to eliminate these effects. The flow cell was optically connected to the spectrophotometer (7) and light source (6) by two optical fibers of 12 mm and 1 m, respectively. All elements were integrated with a software that allowed the control of the many parameters in real time. After systematic analysis and calibration protocols for the UV-vis absorbance experiments, we choose to decrease the signal-noise ratio using a Boxcar Width (≈3200 ms) as integration time and 3 scans to average. MB solutions from 0.08 to 0.8 ppm were flowed to obtain a calibration curve, and the linear fit yielded a final function $y = 2.29 \cdot x$, where $y$ is the absorbance and $x = [MB]$ is the MB concentration.

In all MB degradation experiments, samples were composed by four circular MNFs membranes stacked over each other, providing a total of 3 mg of material, in order to improve the heating response. The four samples were supported on a Nylon filter (0.1 μm pore size, ϕ = 13 mm) to retain potential sample debris that could affect the spectrophotometric measurements by crossing the UV-Vis flow cell. Before each experiment a MB solution ([MB]=0.64 ppm) was circulated during 1.5 h reach adsorption saturation. After this time the MNFs were stimulated turning on the AMF (f = 317.80 kHz; $H_0$ = 28.3 kA·m$^{-1}$) in the D5 equipment to increase the temperature. The MB concentration was monitored continuously in a closed-circuit configuration detecting the 665 nm absorption band and compared with calibration line for the quantification of degraded MB.

## 3. RESULTS AND DISCUSSION

### 3.1 Physical and chemical characterization

The Scanning Electron Microscopy (SEM) image analysis showed that the resulting fibers were quite homogeneous (Figure 2 a) with a narrow distribution of fiber diameter with average value $\phi = 760 \pm 150$ nm, as obtained from the histogram using secondary electrons. It was also found that the MNPs were successfully embedded in the polymeric MNFs (Figure 2 b), with a large amount of MNP within MNFs and on the surface. The nanofiber cross sections obtained from dual beam observation (Figure 2 c) reveals the presence of MNPs embedded within PAN nanofibers. The observed size of the embedded MNPs (40 ± 7 nm) shows good agreement with the 38 ± 6 nm previously reported using similar strategy for MNPs synthesis. [40] As discussed below, the degree of agglomeration did not produce noticeable changes in the heating efficiency of the MNFs compared to the *as prepared* MNPs.

An optical image of a MNFs membrane is presented in Figure 2 d. The sample is supported on a Nylon filter (0.1 μm pore size, ϕ = 13 mm) used for the experiments. The brownish color of the MNFs is given by the high concentration of MNPs in the filaments. Further characterization was made employing the same samples geometries as presented in Figure2 d as commented bellow.

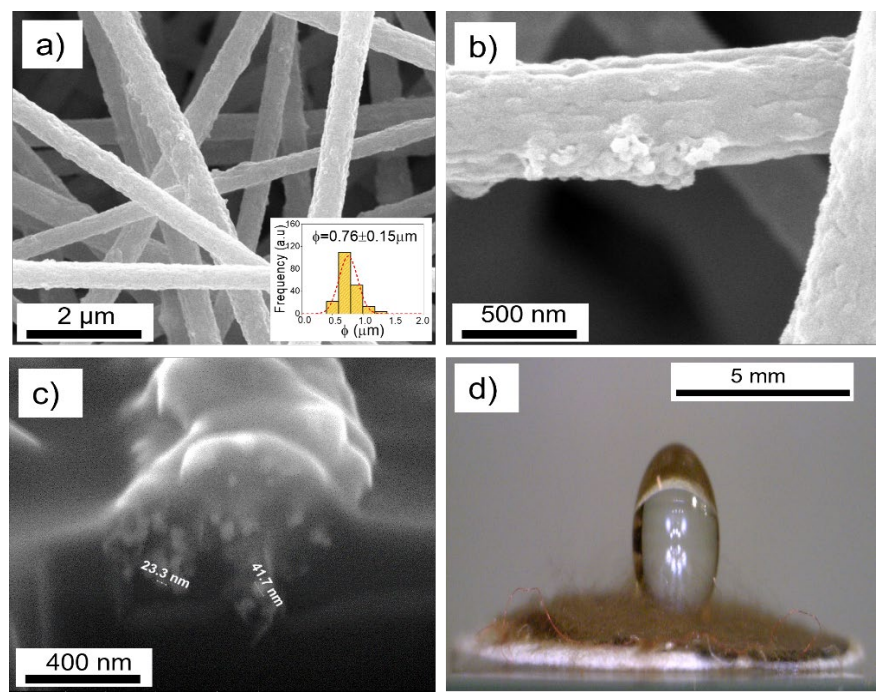

**Figure 2** Scanning electron microscopy images of fabricated magnetic fibers: a) secondary electrons and fiber diameter histogram (inset), b) close-up showing magnetic nanoparticles in the surface, c) cross section of single magnetic fiber, d) optical image of MNFs composed membranes supported on a Nylon filter ($\phi$ = 13 mm).

The Fe and Mn stoichiometry of the MNPs given in atomic percentage (see Figure S1 in the Supplementary Material) showed by EDS analysis was consistent with the spinel ferrite $MnFe_2O_4$ composition. The relationship 80-20 %wt of sulphate precursors ($FeSO_4$ - $MnSO_4$) used during the synthesis resulted in a stoichiometry of $Fe_{2.5}Mn_{0.5}O_4$ (Fe = 83 %, Mn = 17 %) of magnetic material within the MNFs. A minor deficiency of Mn contents was systematically found, as compared with the nominal composition of initial reactants (i.e., ≈ 3 % wt. deficiency from the $MnFe_2O_4$ composition). A standard deviation of SD = 1.43 % was obtained from measurements in three different sample areas (Figure S1), indicating that the sonochemical method produced materials with homogeneous composition that can be supported on electrospun nanofibers with reduced composition modification during the elaboration process.

The resulting MNFs provided a superhydrophobic surface measured by the water-contact angle (WCA) of $\theta_C$ = 165 º (see Figure S2), as expected for PAN nanofibers.[41] The persistence of the superhydrophobic properties after embedding the MNPs indicate a strong chemical interaction between MNPs and the polymer structure, confirmed by the FTIR analysis (Figure S3). A strong absorbance in the visible region (380 - 780 nm) from the MNFs suggest photo-excitability with $E_g$ = 1.75 eV (Figure S4). The thermal stability of MNFs was studied through differential scanning calorimetry measurements (Figure S5 and S6). The data showed a glass transition expected for the pure PAN fibers at a temperatures $T_g$ = 106 °C (Figure S5), which shifted to a value of $T_g$ = 112 °C in MNFs samples. This increase of ≈ 6 °C is attributed to the incorporation of the MNPs into the PAN matrix and confirmed some degree of interaction between PAN chains and MNPs, probably due to the interfacial interaction and possible covalent bonding between PAN and the MNPs surface, reducing the mobility of polymer chains. [42]

The exothermic transition around T ≈ 314 ºC in the PAN fibers corresponds to the oligomerization of the nitrile groups within the PAN structure. This transition was not observed in the corresponding thermogram of MNFs, suggesting that the oligomerization of PAN occurs partially with Nitrogen loses and the characteristic transition of nitrile do not occur completely. PAN-sulfur composites have been reported to shift to lower transition temperatures in function of the sulfur contents. The changes in the thermograms can be explained by the interaction of the carbon chains with the sulfur present in the MNPs, that is pyrolyzed at lower temperature 180 ºC. These thermal analysis results agreed with FT-IR characterization and together show the structural modification of PAN when is mixed with MNPs for MNFs processing. Both analyses support the idea of a more stable polymer after the spinning process with MNPs, related to the residual sulfate contents from the MNPs synthesis.

Magnetic properties revealed (Table S1) that the saturation magnetization of the composite MNFs ($M_S$ = 15 emu·g$^{-1}$ at 10 K and $M_S$ = 9 emu·g$^{-1}$ at 300 K) was lowered as compared to the value measured for the MNPs ($M_S$=53.5 emu/g) as expected. Taking the $M_S$ (T = 10K) this value is consistent with a ≈ 28 wt % of MNPs within the fibers. From the M (H) hysteresis loops (Figure S7) the coercive fields $H_C$ (265 Oe, 10K and 100 Oe, 300K) were consistent with the expected values for $MnFe_2O_4$ spinel. The elaborated magnetic materials showed promising Néel relaxation associated inductive heating efficiency under AMF, as revealed the data showed in Figure S8, where the SLP values were evaluated as a function of applied field (12.4 ≤ $H_0$ ≤ 36.3 kA·m$^{-1}$, f = 323.75 kHz) for MNPs fixed using gelatin, while MNFs were measured using water as a liquid carrier to temperature sensing from the sample (Figure S8). More detailed discussion of these results will be presented in the following section. Moreover, as a catalyst, the MNFs have remarkable potential for their use as water treatment materials thanks to the surface with the presence of $Fe^{2+}$ and $Fe^{3+}$ oxidation states suitable for Fenton-like reactions and $Mn^{2+}$, $Mn^{3+}$, and $Mn^{4+}$ for oxidation of organic compounds in presence of $H_2O_2$ (Figure S9).

## 3.2 Inductive Heating efficiency

The obtained SLP values shown a dependance of $SLP_{(H)} = \Phi \cdot H^\lambda$, with λ= 3.2 and 2.9 for MNPs in gelatin and MNFs, respectively (Figure S8). These values of λ differ from the λ = 2 expected for the LRT approximation. [43] This behavior has been described in previous works assuming that λ > 2 resulting that the area of the M-H hysteresis loop increases in a non-linear manner. [44, 45]

The obtained SLP values from MNPs samples was larger than for MNFs samples at the same AMF. For the maximum field applied, H = 36.3 kA·m$^{-1}$, the SLP values were 139.6 ± 6.4 and 83.7 ± 5.5 W·g$^{-1}$ for MNPs in gelatin and MNFs, respectively, considering the same mass in the samples (3 mg). However, according to the TGA characterization (Figure S 5), about 55% of mass in the MNFs is related to MNPs contents and only the amount of magnetic material in the sample contributes to heating. Assuming a 50% of MNPs contents (discarding 5 % from ashes and residua from non-pyrolyzed polymer) in the MNFs samples, the observed SLP for MNFs is increased (13.8 W·g$^{-1}$), if it is compared with the 50% of the value of MNPs samples, suggesting optimized heating efficiency in the MNFs configuration through agglomeration effects or linear arrangements within the fibers, that were reported as an option for increasing the SLP values for magnetic materials [46-48]. Regarding the heating response of a single MNFs membrane placed in the D5 system under AMF (f = 328 kHz; $H_0$ = 36.6 kA·m$^{-1}$) a temperature of 46 ºC was reached within the first 60 s after the field was turned on (Figure 4 a) at air conditions, which constituted a quite fast thermal response for a mass of MNFs < 1 mg. The maximum temperature that MNFs can be reached is in function of their SLP values and can be modulated by the intensity of the applied magnetic field (H) as can be observed in Figure 3. This response of MNFs can be exploited as remote-controlled local heating agent, for instance, in therapeutic transdermal related therapies. Application of local heating has demonstrated enhancement in the capacity of drugs to crossing the skin barrier [49]. Moreover, against neglected skin infectious diseases the application of local heating is an adjuvant capable to reducing the microorganism population and ensuring sepsis conditions for improved wound healing [50, 51].

The heating response of the MNFs in liquid media were evaluated at acidic, neutral and alkaline pH conditions, flowing solutions sequentially across the samples (Figure 3 b) and stimulated using AMF pulses of 10 min for heating response. Additionally, in a single experiment, different solutions of pH 7, pH 3, and pH 10 flowed through the system (pH mix in Figure 3 c). The maximum reached temperatures were 25.7 ± 0.47, 25 ± 0.45, 25.3 ± 0.5, 25.3 ± 0.27 ºC respectively, showing stability in the heating capacity. Additionally, we tracked the stability of these response against time by repeating the heating experiments at different pH along 25 days after the first experiment. The resulting values of temperature showing negligible variations within this time lapse. However, as shown in Figure 3 c, the evolution of $\frac{dT}{dt}_{max}$ has a decreasing tendency compared with the first measurement. This decrement in the rate after the first cycle can be attributed to magnetic materials losses during the experiment. Is possible that the liquid flux carry with the MNPs attached to the surface, besides to the dissolving effect of acid and basic conditions from MNFs. However, after the second cycle the heating response remains stable and the maximum temperature can be reached. It can be attributed to the response of strongly attached MNPs within the MNFs. The observed performance on heating efficiency from MNFs positioning them as strong candidates for stable inductive heating applications, even in extreme environments as instance, those related to hazardous wastes from industrial activities to the aqueous media.

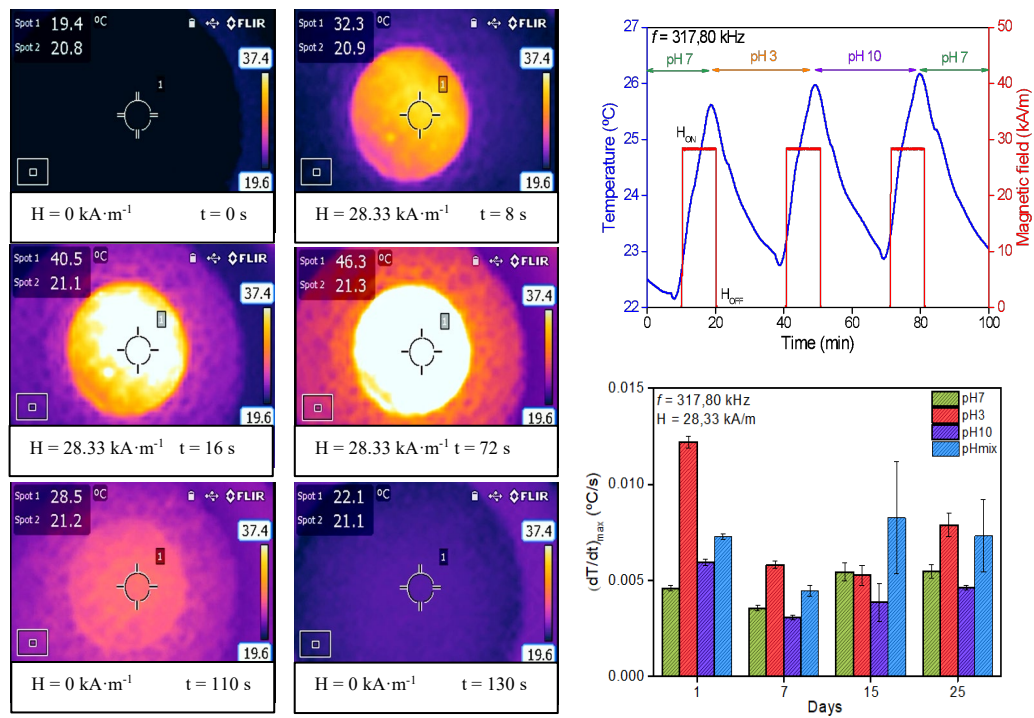

**Figure 3** Heating efficiency of MNFs. a) Thermal images corresponding to the fiber placed in the AMF ($f$ =328 kHz and $H_0$= 36.6 kA·m$^{-1}$) for 60 s; b) heating capacity of MNFs at dynamic flow conditions applying AMF

(317.8 kHz ad 28.3 kA·m$^{-1}$) for 10 minutes. The blue line is temperature (left axe) and the red line the magnetic field applied (right axe), and c) tracking of the heating capacity of fibers with time (*f*= 317.8 kHz; H$_0$=28.3 kA·m$^{-1}$) under aqueous media with different pH values.

## 3.3 Magnetically activated MB degradation

There is a considerable body of experimental and theoretical works on the use of metal and metal oxide nanoparticles on catalytic degradation of organic pollutants in water by enhanced oxidation mechanisms.[52] Those based on iron-containing nanoparticles as active sites for the generation of reactive oxygen species (ROS) have shown remarkable efficacy in the catalytic oxidation of persistent water organic pollutants. [53] However, only some works have reported on the use of magnetic activation to improve or control the catalytic efficacy of magnetic catalysts.[54] Our assessment of MNFs as magnetically-activated catalytic membranes was assessed using methylene blue (MB) as a target molecule.

Before the experiment, 5 mL of a solution of 0.64 ppm of MB with H$_2$O$_2$ (1.28 μL) was flowed (Q = 0.4 mL·min$^{-1}$) for 1.5 h to reach equilibrium and possible adsorption of MB on the MNFs. Once the concentration was stable for several minutes in the UV-Vis module (Figure 1 c), the magnetic field was turned on and the MNFs were under AMF (H$_0$ = 24.8 kAm$^{-1}$; *f* = 317,18 kHz) while the concentration of MB was monitored during 4 hours. Four circular MNFs membranes (ϕ = 10 mm) stacked together were employed and a target temperature T=35ºC was set during magnetic heating. The control of the target temperature was achieved by a proportional–integral controller (PI) using H$_0$ as feedback parameter (See Figure 4, lower panel).

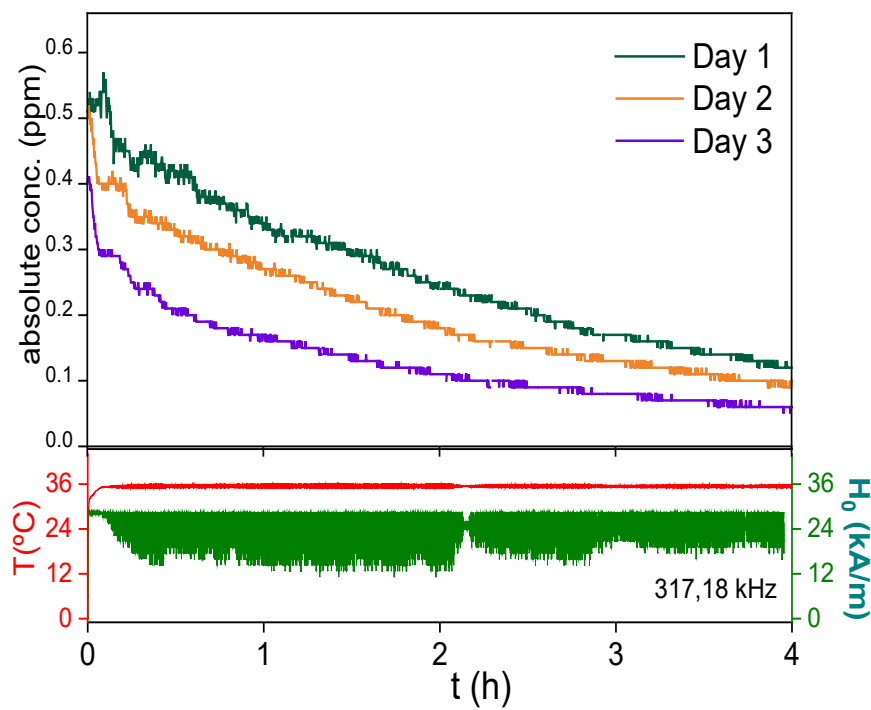

**Figure 4**. Upper panel: time dependence data of MB concentration during degradation by MNFs triggered by AMF ($H_0$=24.8 kAm$^{-1}$; $f$=317,18 kHz), in the presence of $H_2O_2$ (1.28 ppm) at pH 7. The three curves replicate the experiment re-using the same MNF sample along three different days. Lower panel: registry data for the set point temperature (35ºC, red line) and the feedback response of $H_0$ amplitude (green line and right axe) controlling the temperature by a PI software with fast changes in $H_0$ ($\Delta t \approx 5$ s). Note that the fast changes in $H_0$ cannot be resolved in the plotted scale, producing the appearance of a 'thickening' in the $H_0$ (green) curve.

The reusability of the MNFs membranes for MB removal was evaluated three times, one time a day, drying the membranes over night at 60 ºC, air atmosphere. For each experiment, initially the MNFs were put in contact at 35 ºC with a circulating pure water and kept it for ≈30 min to wet the MNFs. After, the medium was replaced instantly, changing the circulating path (using a zero-dead volume switch valve) from water to the MB solution. The catalytic activity of MNFs was measured through the decrease of MB concentration over the time. The measured degradation time dependence (Figure 4) was fitted using the Langmuir–Hinshelwood (LH) model [55] [56] [57], developed for heterogeneous photocatalysis through the expression [58]:

$$\frac{1}{r} = \frac{1}{k_r} + \frac{1}{k_r \cdot K \cdot C} \qquad (Eq.\ 4)$$

where $r$ is the rate of the reaction that changes with time, $k_r$ is the limiting rate constant of reaction, $K$ is the equilibrium constant for adsorption of the substrate onto catalyst and $C$ is the concentration at any time. This model includes the reversible binding of substrate molecules with a multiplicity of active sites at the nanoparticle surface, assuming fast adsorption mechanisms that provides equilibrium at all times.[59, 60] The LH model´s goodness of the fit to the experimental data was evaluated using the Nash–Sutcliffe model efficiency (NSE) coefficient criterion as hypothesis testing indicator, using root mean square error (RMSE) and standard deviation (SD) as follows $NSE = 1 - (\frac{RSME}{SD})^2$. This coefficient of efficiency of mathematical predictions has a perfect fit at NSE = 1 [61]. Table1 summarizes the kinetics parameters such as degradation percentage ($\eta$), using the relation $\eta = \frac{[MB]_0 - [MB]}{[MB]_0} \times 100$, total mass degraded ($m_t$) and the LH degradation kinetics constants $k_r$ and $K$, and goodness-of-fitting according to NSE.

**Table 1** Degradation percentage ($\eta$), total mass degraded ($m_t$), Langmuir - Hinshelwood constants for magnetically activated degradation of MB at T = 35 °C, and goodness-of-fitting according to Nash–Sutcliffe model efficiency (NSE) coefficient criterion.

|        | $\eta$ [%] | $m_t$ [µg] | $k_r$  | K      | RSME ·10$^{-4}$ | SD$_{exp}$ | SD$_{model}$ | NSE    |
|--------|------------|------------|--------|--------|-----------------|------------|--------------|--------|
| Test 1 | 81.7       | 2.4        | 356.52 | 0.0028 | 1.54            | 0.1105     | 0.1105       | 0.9999 |
| Test 2 | 85         | 2.5        | 356.52 | 0.0028 | 1.47            | 0.0928     | 0.0927       | 0.9999 |
| Test 3 | 91.7       | 2.7        | 355.58 | 0.0028 | 0.87            | 0.0669     | 0.0669       | 0.9999 |

The obtained value of $\eta > 80\%$ since de first cycle of usage show adequate capabilities for the implementation of MNFs as degradation agent driven by AMF with any decrement on their capacity after three experiments. Also, the degraded mass is similar at each case, ensuring stability and reusability of the MNFs as degradation agent. Calculated values of $k_r$ and $K$ were the same in the three cases, this behavior can be associated to stable values for adsorption equilibrium and limiting rates. We can stablish those values as characteristic from the catalytic activity of MNPs embedded in nanofibers of PAN, reported for first time in this work, thanks to the goodness-of-fitting (NSE = 0.9999). As ca be observed, the SD values are reduced and quite similar between SD$_{exp}$ and SD$_{model}$, ensuring that the experimental data are well forecasted by the LH heterogeneous catalysis kinetics. These results were consistent with a previous work [62] where adsorptions and discoloration of MB was negligible at all temperatures tested due to the MB surface was protonated, except at pH 10 when adsorption was clearly detected. When compared to the works reported by Muñoz et al. [36] and Rivera et al. [63], our experimental conditions reflected and improved degradation efficiency since a lower concentrations of $H_2O_2$ was required for our membranes, and at lower temperatures, for similar catalytic activity and degradation % of MB.

## 4. CONCLUSIONS

The synthesis protocol presented in this work to prepare multifunctional MNFs allowed us to obtain polymeric/magnetic composite membranes capable of remotely triggered catalytic reactions, thanks to the embedded MNPs acting as both local nanoheaters and catalysts. This novel composite showed chemical stability against extreme acidic and basic media, while its response to a.c. magnetic fields provided a way to remotely switch the catalytic activity on and off thus controlling the redox mechanisms. The degradation kinetics for MB could be explained in terms of a Langmuir–Hinshelwood mechanism for heterogeneous catalysis, in which the degradation rate is limited by the reaction constant with a slight contribution of the adsorption of molecules on the active sites, with some features resembling allosteric enzymatic activity. Overall, these results showed the potential of MNFs and hydrophobic membranes as removal agents for contaminants in fluxing water, e.g., underground pipelines with limited access. Although we could demonstrate the working principle on these materials and industrial-scale translation seem plausible, further optimization of the degradation efficiency will be needed for environmental applications such as biological wastes or microplastic degradation.

# Supplementary Information

## 1. EDS analysis

Compositional analysis from MNFs were performed using Energy Dispersive Spectroscopy (EDS). The spectra were obtained at 20 kV acceleration voltage for adequate excitation of X-Ray photons from Fe ($L_\alpha$ = 0.705 keV) and Mn ($K_\alpha$ = 5.894 keV) elements and the relationship among elements were determined from the tables provided by the equipment´s software (INCA Energy® Software, Oxford Instruments®). Figure S1 shows SEM images (a) where the selected analysis area is denoted by pink squares and the respective table can be observed under each image (b). The table with the calculated average amount of Fe and Mn (c) shows the Mn relationship in the material and a proposed stoichiometry is provided in a $Fe_{3-x}Mn_xO_4$ structure.

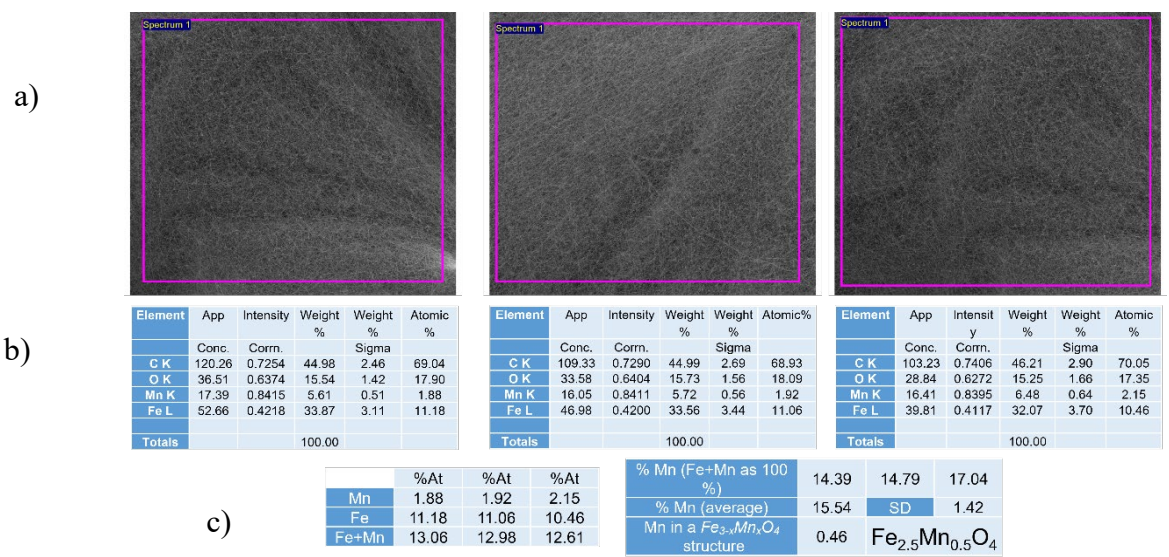

## 2. Contact angle

The contact angle of the membrane´s surface composed by MNFs was determined using a water drop (10 µL) deposited on its surface, then a photo was taken and the contact angle was determined using image processing software. MNFs membrane reveal contact angle of 165º, corresponding to a superhydrophobic behaviour (contact angle value > 120º) as can be observed in the image of Figure S2.

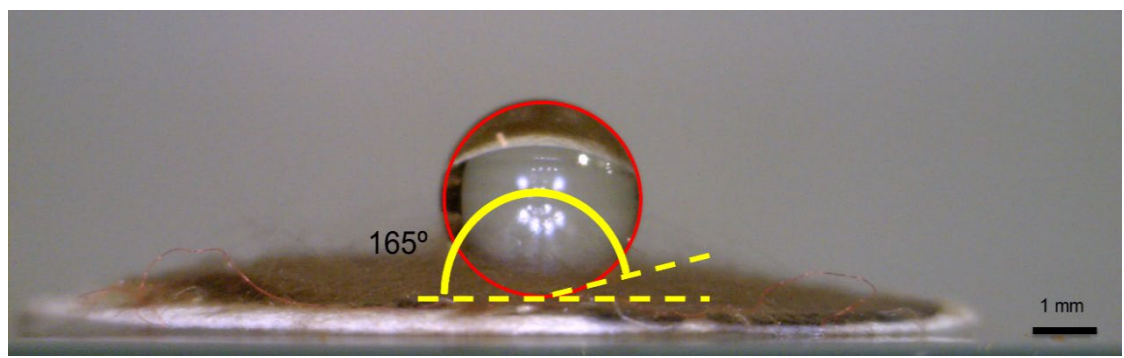

## 3. FT-IR

In the Figure S3, FTIR-ATR spectra of the PANFs (black) are compared with MNFs (red) to observe if the processing of polymer with MNPs and the electrospinning process produces conformational changes on the polymeric structure or detecting chemical interactions among the components. The bands from normal vibrations of functional groups present in the PANFs sample reveal amine vibrations were manifested from 3000 – 3600 cm$^{-1}$ and typical vibrations from $v_{as}$-CH$_2$- skeleton of PAN at 2930 cm$^{-1}$ were detected. Also, on this band are contained the vibrations from carbonyl groups: Fermi resonance of aldehyde, confirmed by the band centered at 2640 cm$^{-1}$, this bands together with the stretching of OH in carboxylic groups (2321 cm$^{-1}$) are related to the PAN dissolution in DMF. Characteristic stretching vibrations of C≡N (2247 cm$^{-1}$) and those attributed to -CO-CRN$_2$ at 2067 cm$^{-1}$ were identified. Double bonds (C=C, C=N, C=O, and C–C=C–N promotes the apparition of the band centered at 1660 cm$^{-1}$, also related with the DMF processing. 1450 cm$^{-1}$ band reveals the CH$_2$ bending (scissoring) in PAN and 1364 cm$^{-1}$ the C-H interactions of the main chain. The vibration of C=N (1233 cm$^{-1}$), and =CH (807 cm$^{-1}$) respectively appears in the spectrum. Finally, primary and secondary amines interactions were detected from 900 to 650 cm$^{-1}$ due to DMF interactions with PAN.
In the case of MNFs, main vibrations of PAN chains were observed, with some spectral changes and new bands appears as effect of the MNPs addition to the solution during their elaboration. Amine vibrations (3000 – 3600 cm$^{-1}$) were masked by a broad band centered at 3320 cm$^{-1}$ related to S-H bonds, correlated with 1224 cm$^{-1}$ from C-S, -SO$_3$ (1050 cm$^{-1}$), S–S (939, 515 cm$^{-1}$) and C–S (669 cm$^{-1}$), revealing remaining contents of sulphates in the MNPs powder. While the formation of the small band centered at 1630 cm$^{-1}$ related to C=C, C=N and N-H bonds is associated to dehydrogenation and stabilization of PAN structure. Additionally, some functional groups of C=C (1504 cm$^{-1}$) of the aromatic ring and 1170 cm$^{-1}$ unsaturated esters and carboxylic acids are present.

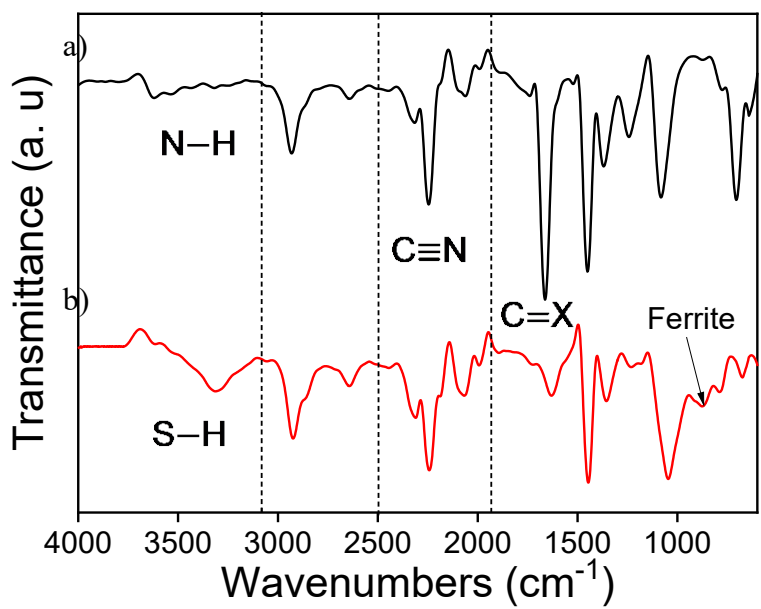

Figure S3. FTIR spectra of samples a) PAN fibers as blank (PANFs) and MNFs b) showing the functional groups vibrations modifications as MNPs are embedded within polymeric chains.

## 4. UV-Vis characterization

The absorbance band centered at 268.5 nm from PANFs sample in the UV-vis spectrum shown in Figure S4 (black line) is attributed to cyanoenamine structure that absorbs around 265 nm, also this band is in good agreement with reported spectral behavior from PAN nanofibers processed using DMF as solvent [1, 2]. The Tauc´s plots Figure S 4 b-c reveal band gap energy values ($E_g$) that positioning to the PANFs as insulator ($E_g$=5.72 eV), while the MNFs sample shown a reduced $E_g$ value (1.75 eV), this effect can be associated to the presence of well-distributed MNPs within the polymeric fibers. The showed $E_g$ value is close to other values related with FeMnO compounds reported elsewhere [3] [4] indicating a direct allowed transition (n=2 in Tauc´s plot). A broad band centered at 413 nm in the UV-vis spectrum (Figure S4 red line) manifest strong absorbance of MNFs sample in the visible region (380 - 780 nm).

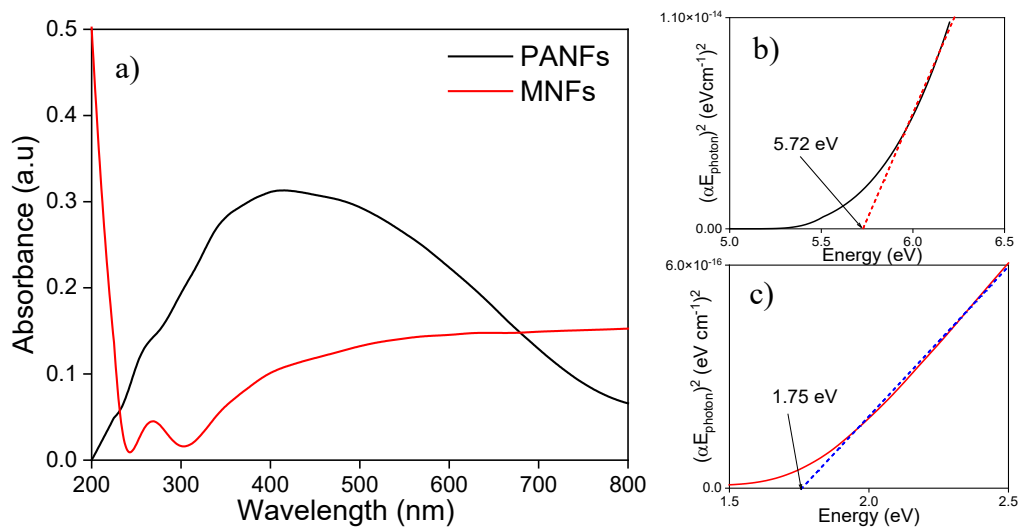

Figure S4  UV-vis spectra from 200 to 800 nm (a) of PANFs (black) and MNFs (red). Tauc´s plots in the right panels show the linear adjust for the determination of the band-gap energy, in the case of PANFs (b) corresponds to an insulator, while MNFs sample (c) shows direct allowed transitions.

## 5. Thermal analysis

### 5.1 Thermogravimetric analysis

Thermogravimetric curves of PANFs and MNFs can be observed in Figure S5. The curves shown the weight loss in function of the pyrolytic reactions occurred from 50 to 800 ºC for PANFs and MNFs (Figure S5 a) and b), respectively). In the case of MNFs, possible evaporation of solvent (DMF) and humidity produce weight loss (1.2 %) around 110 ºC. Volatile gases evolved in the cyclization of PAN represent the 20 % of weight loss around 316 ºC and the 75 % of weight loss around 552 ºC is associated to the evaporation of ammonia, HCN and oligomeric products. The complete evaporation of PANFs starts at 600 ºC. On the other hand, MNFs shown 17 % of weight loss at increased temperature of transition (330 ºC) associated to good intermolecular cohesion between MNPs and polymeric chains and reduced cyclization transition. In this sample, the degradation starts at lower temperature and the 35 % of loosed weight occurs around 448 ºC. The complete evaporation of polymer starts at 470 ºC and a constant 55% of residue is attributed to the MNPs contents that are not volatile in this temperature interval.

a) 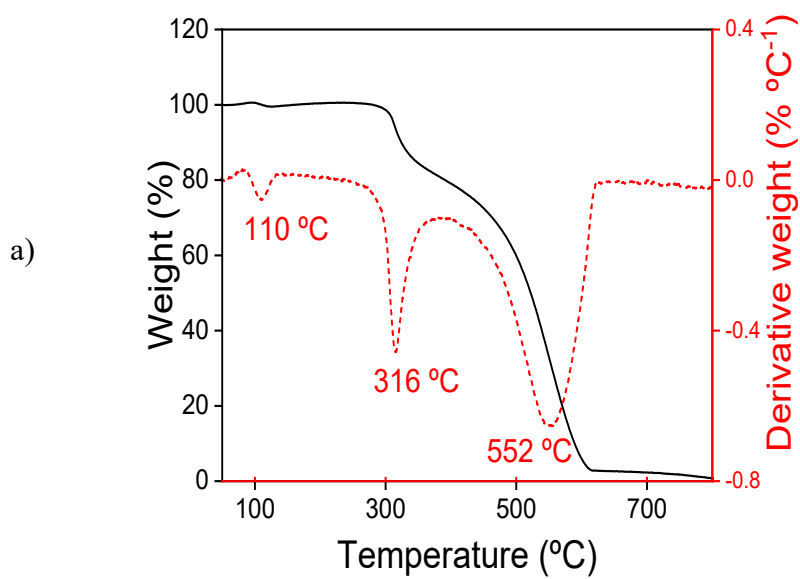

b) 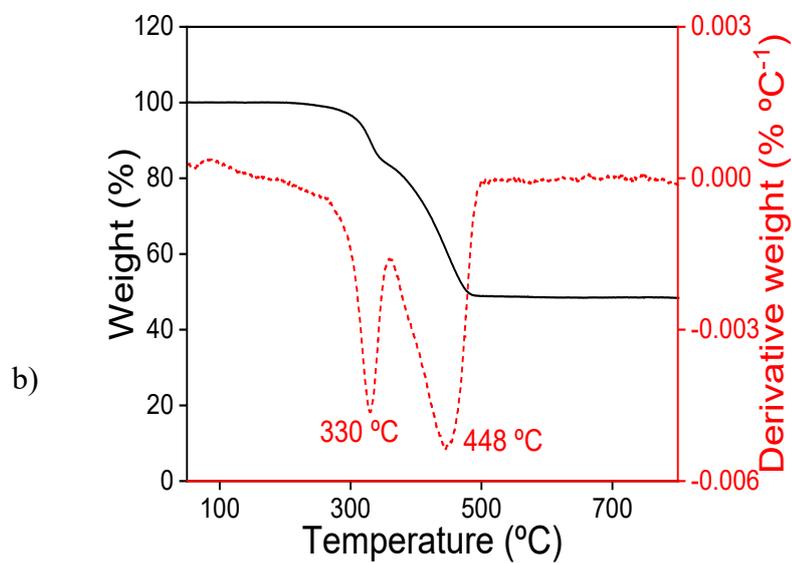

Figure S5. Thermograms of samples showing weight losses (black) and derivative weight losses (dashed red): a) PANFs and b) MNFs

## 5.2 Differential scanning calorimetry

Differential scanning calorimetry curves showed the glass transitions temperatures ($T_g$) of PANFs and MNFs samples, as illustrated in Figure S6. It can be seen that, compared with PAN nanofiber ($T_g$ = 106 °C), the $T_g$ was increased by about 6 °C by incorporating MNPs into the PAN matrix ($T_g$ = 112 °C). The observed increase in $T_g$ as compared to the PANFs confirms the interaction between PAN chains and MNPs. The improvement in the $T_g$ stemmed from a stronger interfacial interaction and possible covalent bonding between PAN and the MNPs. The results suggested that the mobility of PAN chains was reduced due to the constraint effect of MNPs.[5]

The exothermic transition around 314 ºC in the PANFs thermogram corresponds to nitrile groups oligomerization in the PAN structure, that is not present in the corresponding thermogram of MNFs, suggesting that the oligomerization of PAN occurs partially with Nitrogen loses and the characteristic transition of nitrile do not occur completely. PAN-sulfur composites have been shown shifts at lower transition temperatures in function of the sulfur contents. The changes in the thermograms can be explained by the interaction of the carbon chains with the sulfur present in the MNPs, that is pyrolyzed at lower temperature 180 ºC. These thermal analysis results agreed with FT-IR characterization and together show the structural modification of PAN when is mixed with MNPs for MNFs processing. Both analyses reveal results associated to stabilization of polymer after the spinning process in the MNFs related with the remaining sulfate of MNPs synthesis.

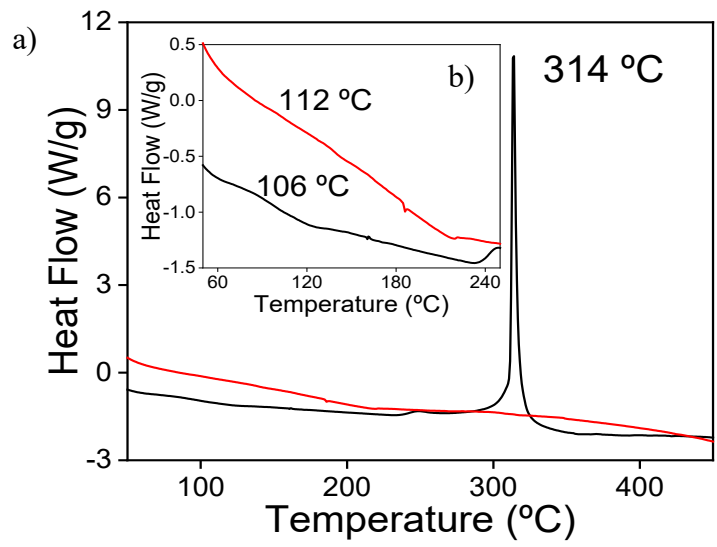

Figure S6. Thermograms from DSC technique from 50 to 600 ºC (a) of PANFs (black) and MNFs (red) as insets (b) 50 to 250 ºC interval showing the glass transitions temperatures (Tg) shift induced by the MNPs within polymeric fibers

## 6. Magnetic characterization

The magnetic properties of MNFs such as saturation magnetization ($M_s$) and coercive field ($H_c$) were determined analysing the hysteresis curves (M *vs* H, Figure S7) at 10 and 300 K and reported in Table 1.

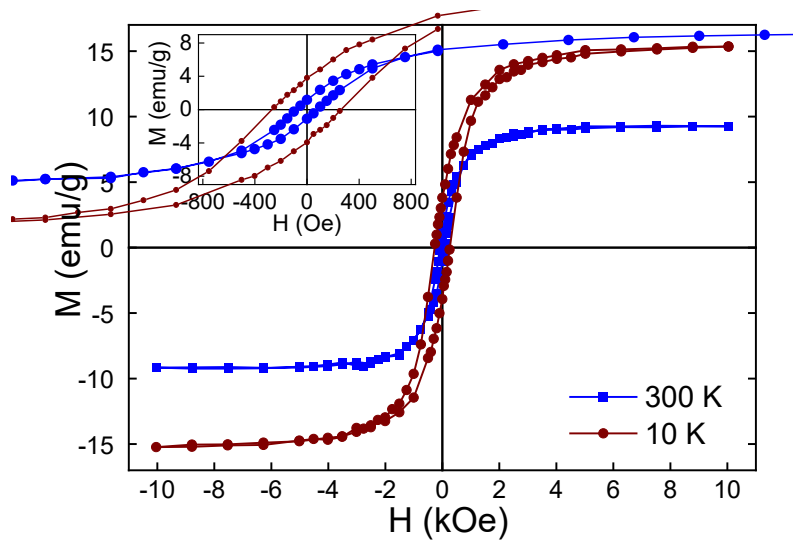

Figure S7. Magnetitic behaviour of MNFs measured at 10 and 300 K, as inset, a zoom in the -800 to 800 Oe interval is presented showing the coercive field $H_c$.

## 7. Specific loss power from MNPs and MNFs

The behaviour of the obtained SLP values as a function of applied field ($12.4 \leq H0 \leq 36.3$ kA·m-1, f = 323.75 kHz) using MNPs fixed using gelatin and MNFs were measured using water as a liquid carrier can be observed in Figure S8. The experimental points were fitted to a power dependence of $SLP_{(H)} = \Phi \cdot H^\lambda$, with $\lambda$= 3.2 and 2.9 for MNPs in gelatin and MNFs, respectively.

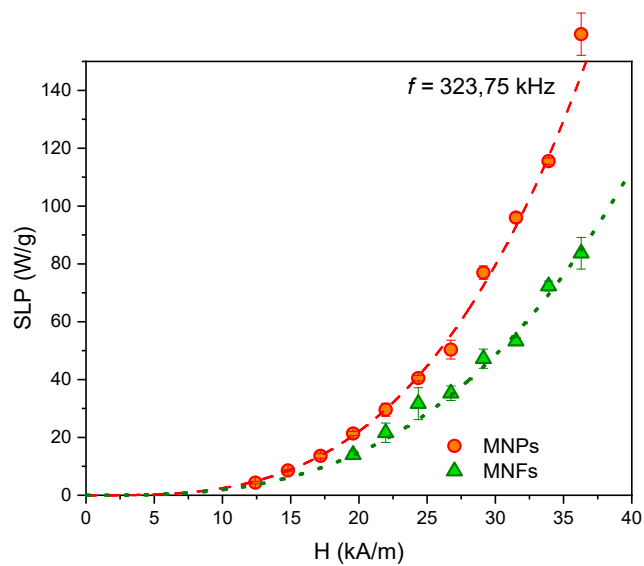

**Figure S8** Dependence SLP vs. H on magnetic field for MNPs (green triangles) in gelatin and MNFs (red circles) in water. Dashed lines represent the power fitted curves SLP(H) = $\Phi \cdot H^\lambda$. In both curves, data in the vertical axis is given in Watts per gram of MNPs.

## 8. X-Ray photon spectroscopy analysis

Figure S9 a) shows data about XPS analysis in the 1200 to 0 eV binding energies interval. Signals from N and O KLL, O 1s, N 1s and C1s are predominant in the surface of MNFs and the interactions from Fe and Mn elements cannot be resolved. However, high-resolution observations reveal signals related to those elements as can be observed in Figure S 9 b). From 740-700 eV (Figure S 9 b)) the Fe 2p signal splits in 723 eV (Fe $2p_{1/2}$) and 710 eV (Fe $2p_{3/2}$) bands together with the presence of satellite peak centred at 718 eV from $Fe^{3+}$. Deconvolution of Fe $2p_{3/2}$ band reveals peaks associated to the presence of $Fe^{2+}$ (12%) and $Fe^{3+}$ (78%) on the MNFs surfaces, while Mn 2p interval from 660 to 635 eV (Figure S 9 c)) shows $Mn^{2+}$ (26%), $Mn^{3+}$ (3.5%), and $Mn^{4+}$ (13.5%) interactions combined with the signal associated to MnO and MnOOH (57%) [6-8].

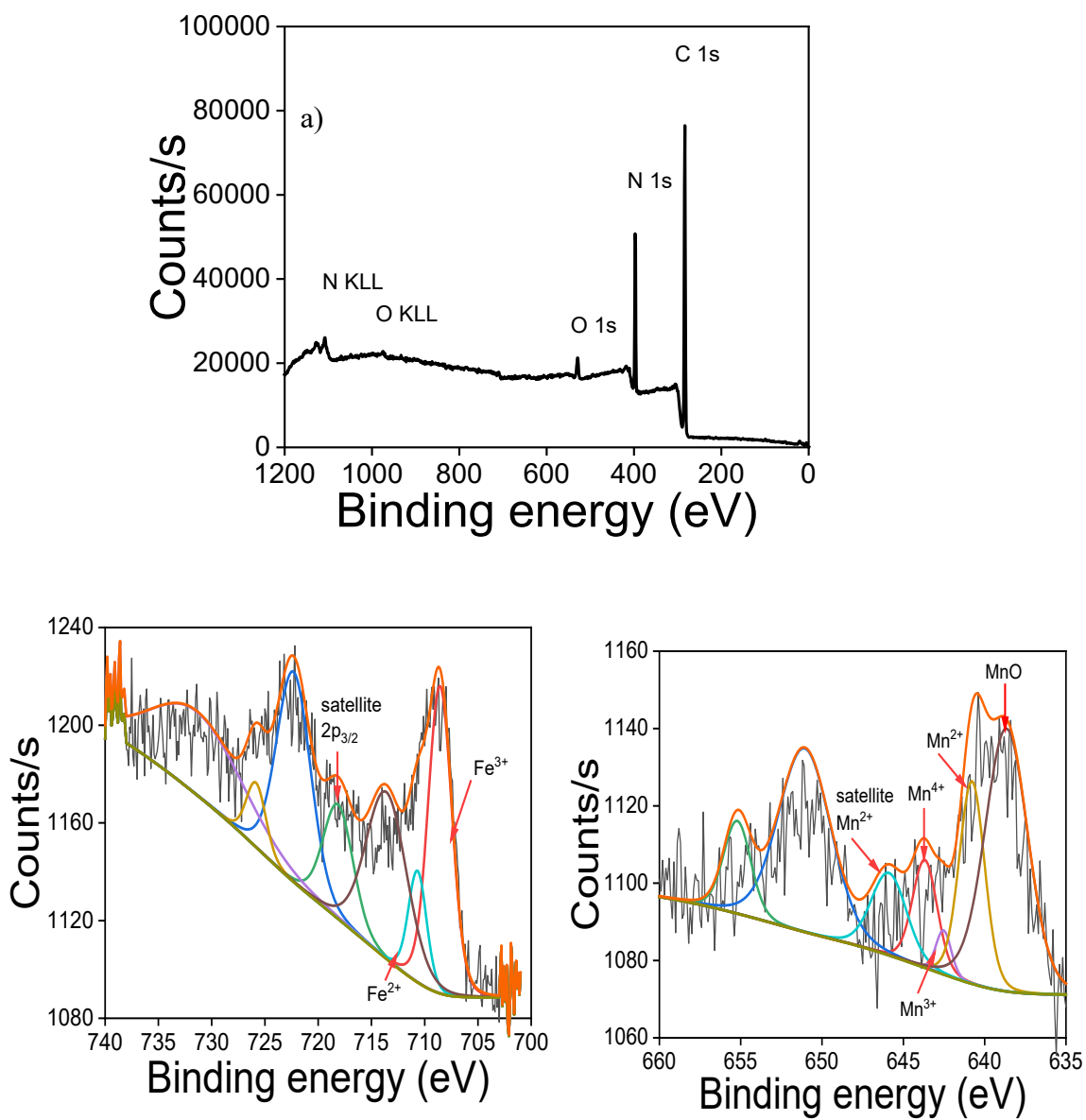

Figure S 9. X-ray photon spectroscopy spectra of MNFs surfaces. The complete survey shows the interactions of N, O and C elements a). High-resolution survey for Fe 2p b) and Mn 2p c) are presented.

## 9. Inductive heating behaviour

Four measurements at different (1, 7, 15 and 25) days were performed to obtain the curves of Figure S10, that shows the temperature increasing produced by MNFs as response to the alternating magnetic field stimulation under different pH conditions. Monotonic response can be observed under the different conditions, probing the capacity of the elaborated material to produce heating in acid, neutral and alkali solutions. Moreover, this capacity for inductive heating is preserved along the time as can be observed in the curves. Slight differences among measurements can be founded and can be associated to the interaction of the solution flux with the MNPs.

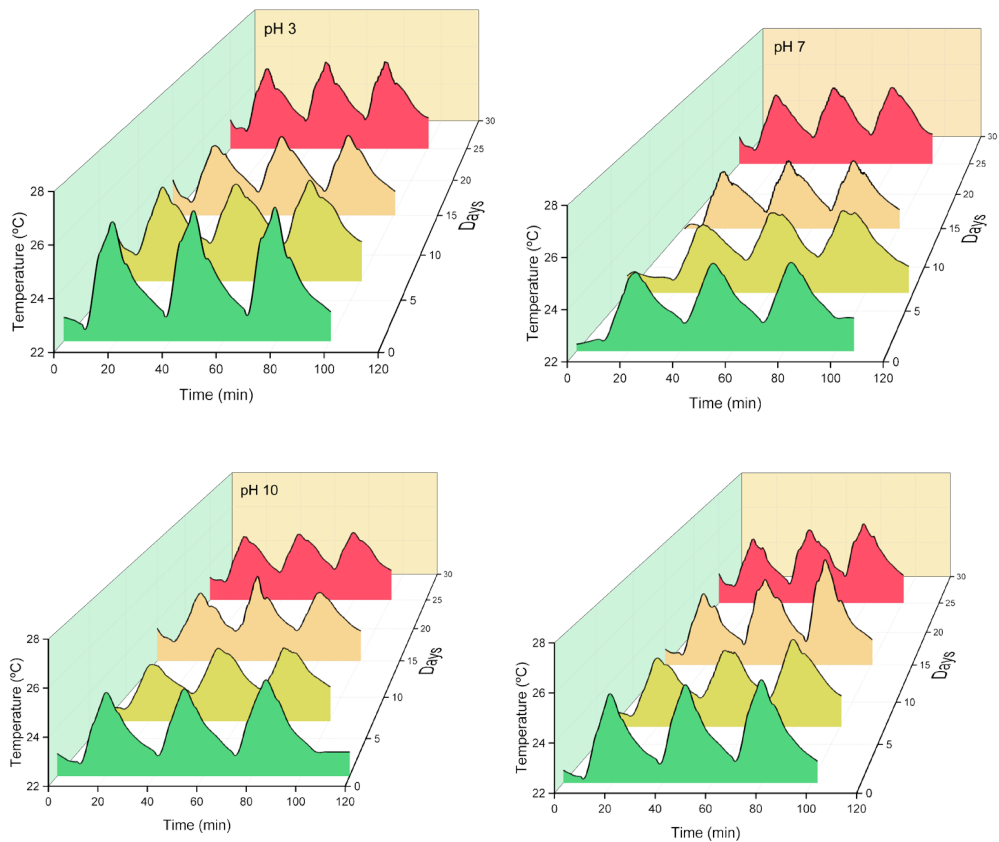

Figure S10: Membranes inductive heating performance in function of time at different pH conditions

## 10. Morphological stability of magnetic nanofibers

MNFs preserves their morphology after the inductive heating behaviour experiments as can be observed in Figure S11. SEM images reveals non-fragmented and homogeneous filaments. Also, MNPs can be observed, probing that they are strongly embedded into the fibers.

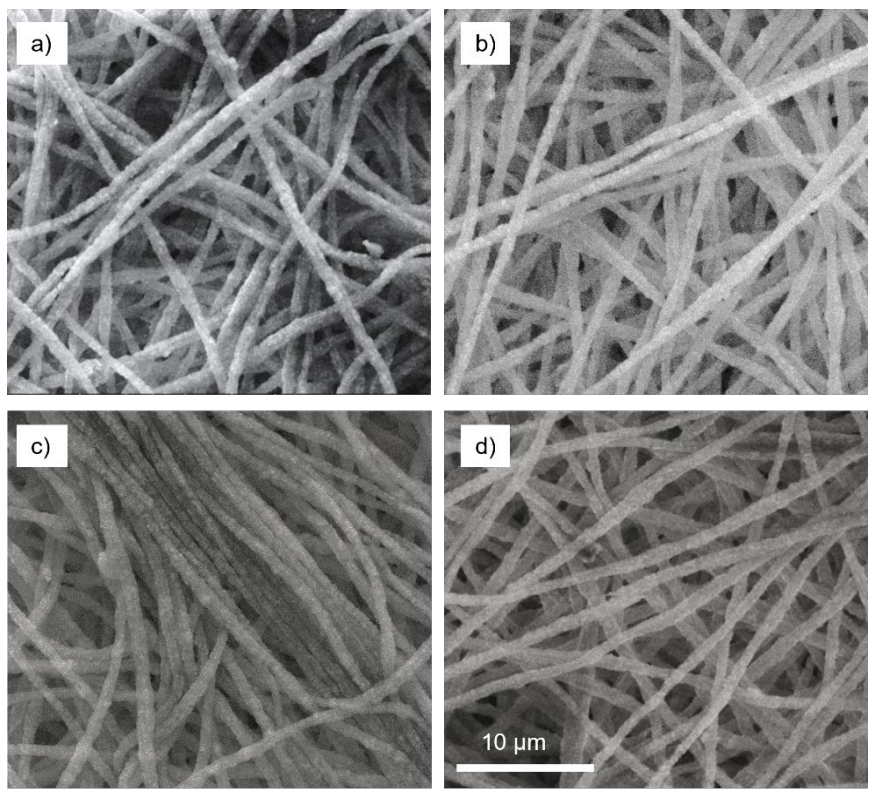

Figure S 71: SEM images of MNFs after inductive heating experiment performed at: a) pH=7, b) pH=3, c) pH=10, and d) pH mix.